\documentclass[pra,showpacs,twocolumn,amsmath,amssymb]{revtex4}
\usepackage{graphicx}
\usepackage[pdfstartview=FitH,bookmarks=false]{hyperref}
\usepackage{bm}

\begin{document}

\title{Temperature driven structural phase transition for trapped ions \\
and its experimental detection}
\author{Zhe-Xuan Gong}
\affiliation{Department of Physics and MCTP, University of Michigan, Ann Arbor, Michigan
48109, USA}
\author{G.-D. Lin}
\affiliation{Department of Physics and MCTP, University of Michigan, Ann Arbor, Michigan
48109, USA}
\author{L.-M. Duan}
\affiliation{Department of Physics and MCTP, University of Michigan, Ann Arbor, Michigan
48109, USA}
\date{\today }

\begin{abstract}
A Wigner crystal formed with trapped ion can undergo structural phase
transition, which is determined only by the mechanical conditions on a
classical level. Instead of this classical result, we show that through
consideration of quantum and thermal fluctuation, a structural phase
transition can be solely driven by change of the system's temperature. We
determine a finite-temperature phase diagram for trapped ions using the
renormalization group method and the path integral formalism, and propose an
experimental scheme to observe the predicted temperature-driven structural
phase transition, which is well within the reach of the current ion trap
technology.
\end{abstract}

\maketitle

Ions trapped in a linear Paul trap or a planar Penning trap have become a
very useful platform \cite{WPW99}, with exciting applications in both
quantum information science \cite{HRB08} and precision measurements \cite%
{BMB98}. Trapped ions also provide a controllable system to simulate and
study many-body phase transitions \cite{FSG08}. A well-known phase
transition that can be observed in a small ion crystal is the structural
phase transition of the Wigner crystal formed with trapped ions, which has
raised significant interest and been extensively studied \cite%
{HKT01,KD03,MNM06,BKW92,WKB91,ESG00,Dubin93,Schiffer93,MF04E,MF04,FCC08,RTS08}. For
instance, a linear crystal in a Paul trap can be squeezed to a zigzag shape
with change of the aspect ratio between the transverse and the axial
trapping frequencies. The structural phase transition for trapped ions so
far is formulated on a classical level, determined by the mechanical
equilibrium conditions. On a classical level, quantum and thermal
fluctuation of the ion positions play no role in the structural phase
transition, and this transition is thus independent of the system's
temperature.

In this paper, we develop a theoretical formalism to take into account
quantum and thermal fluctuation in the structural phase transition, and show
for the first time that a structural phase transition can be driven solely
by change of the system temperature. The structural phase transition is
induced by condensation of phonons into the soft mode (the lowest frequency
collective oscillation mode of the ion crystal). Anharmonic coupling between
different phonon modes intrinsic in the Coulomb interaction leads to
renormalization of the soft mode frequency which affects the phase
transition point. We calculate the system's partition function using the
path integral approach, and gradually integrate out the high frequency modes
with the renormalization group (RG) method to construct the RG flow for the
soft mode frequency. With this formalism, we can calculate the
finite-temperature phase diagram for the ion crystal. Using the linear ion
crystal in a Paul trap as an example, we propose an experimental scheme to
detect the predicted temperature-driven linear-to-zigzag structural phase
transition and show that the requirements in observing this transition fits
well with the current status of the experimental technology.

We consider $N$ ions of mass $m$ subject to external harmonic potentials in
both axial ($z$) and transverse ($x,y$) directions. To be concrete, we take
a linear Paul trap as an example with the trapping frequencies $\omega
_{y}>\omega _{x}>\omega _{z}$ (the method can be extended easily
to other type of traps). We consider the system near the linear-to-zigzag
transition point, with the ions distributed along the $z$ direction with a
tendency towards the zigzag transition in the $x-z$ plane. To describe this
phase transition, it suffices to consider the ion interaction Hamiltonian in
the $x-z$ plane, given by
\begin{equation}  \label{1}
H=\sum_{i=1}^{N}\sum_{\alpha =x,z}[\frac{p_{i\alpha }^{2}}{2m}+\frac{1}{2}%
m\omega _{\alpha }^{2}\alpha _{i}^{2}]+\sum_{i>j}\frac{\kappa }{\left\vert
\mathbf{r}_{i}-\mathbf{r}_{j}\right\vert },
\end{equation}
where $\kappa $ is the Coulomb interaction rate. We assume the temperature
of the system is significantly below the melting temperature of the ion
crystal, which is typically of the order of $0.1-1$K \cite{PSB04}. This
condition is satisfied straightforwardly in experiments with laser cooling.
The ions have well-defined equilibrium positions $\overline{\mathbf{r}}_{i}$%
, and we expand $\mathbf{r}_{i}$ around the equilibrium positions up to the
fourth order of the displacement operators $\delta \mathbf{r}_{i}\equiv
\mathbf{r}_{i}-\overline{\mathbf{r}}_{i}$. Up to the second order of $\delta
\mathbf{r}_{i}$, the quadratic part of the Hamiltonian can be diagonalized
to get the normal phonon modes. For $N$ ions in the $x-z$ plance, there are
in total $2N$ normal modes, and we label them from $1$ to $2N$ in the
ascending order of the mode eigen-frequencies. Expressed with the
coordinates of the normal modes, the Hamiltonian has the form

\begin{eqnarray}
H &=&\sum_{i=1}^{2N}\frac{p_{i}^{2}}{2m}+\frac{1}{2}m\omega _{z}^{2}z_{0}^{2}%
\biggl(\sum_{i=1}^{2N}\omega _{i}^{2}q_{i}^{2}  \nonumber  \label{2} \\
&+&\sum_{ijk}^{2N}B_{ijk}\,q_{i}q_{j}q_{k}+\sum_{ijkl}^{2N}C_{ijkl}%
\,q_{i}q_{j}q_{k}q_{l}\biggr)
\end{eqnarray}%
where $p_{i}$ and $q_{i}$ are the canonical momentum and coordinate for the $%
i^{th}$ phonon modes and $\omega _{i}$ denotes the corresponding
eigen-frequency. We have factorized out $\omega _{z}$ (axial trap frequency)
and $z_{0}\equiv (2\kappa /m\omega _{z}^{2})^{1/3}$ (typical distance
between the ions) as the frequency and the length units ($\omega
_{i},q_{i},B_{ijk},C_{ijkl}$ are thus all dimensionless). The terms with $%
B_{ijk}\,$ and $C_{ijkl}\,$ represent the cubic and quartic terms in the
expansion of the Coulomb potential, and we need to keep both of them as they
lead to the same order of correction to the phase transition point in the
following renormalization calculation. The values for $\omega _{i}$, $%
B_{ijk},\,$and $C_{ijkl}\,$ are determined numerically through expansion of
the Hamiltonian in Eq. (\ref{1}) and diagonalization of its quadratic
components \cite{LZI09}.

The structural phase transition is caused by phonon condensation in the
lowest normal mode (soft mode, or mode 1 in our notation, which corresponds
to the zigzag mode for an ion chain). This happens when the effective
frequency $\omega _{1eff}$ of the soft mode crosses zero. In the classical
treatment \cite{FCC08}, interaction and fluctuation of the phonon modes are
neglected and the effective frequency $\omega _{1eff}$ is just given by the
bare frequency $\omega _{1}$ in the Hamiltonian (\ref{2}). As $\omega _{1}$
is determined simply through expansion and diagonalization of the trapping
and the Coulomb potentials, it is apparently determined only by the
mechanical conditions and has no dependence on the system's temperature.
Here, we take into account the phonon interaction and derive the effective
frequency $\omega _{1eff}$ through a renormalization group treatment of the
partition function corresponding to the Hamiltonian (\ref{2}) in the path
integral formalism. As a qualitatively new result from this treatment, we
show that the structural phase transition is not purely mechanical any more
and becomes a thermodynamic transition depending on the system temperature.

In the path integral formalism, the partition function of the system $%
Z=e^{-H/(k_{B}T)}$ (where $T$ is the system temperature) can be written as
\cite{Kleinert04}
\begin{equation}
Z=\oint \prod_{i=1}^{2N}\mathcal{D}q_{i}e^{-S},  \label{3}
\end{equation}%
where the action
\begin{eqnarray}
S &=&\int_{0}^{\hbar \omega _{z}/(k_{B}T)}\frac{d\tau }{\hbar \omega _{z}}%
\frac{1}{2}m\omega _{z}^{2}z_{0}^{2}\left\{ \sum_{i=1}^{2N}\left[ (\partial
q_{i}/\partial \tau )^{2}+\omega _{i}^{2}q_{i}^{2}\right] \right.   \nonumber
\label{4} \\
&&\left.
+\sum_{ijk}B_{ijk}\,q_{i}q_{j}q_{k}+\sum_{ijkl}C_{ijkl}%
\,q_{i}q_{j}q_{k}q_{l}\right\} .
\end{eqnarray}

The RG method provides a way to work out this partition function and to find
the effective frequency $\omega _{1eff}$ of the lowest mode \cite{PS95}. The
basic idea of the RG method is to integrate out the high frequency modes in
the path integral step by step to get a renormalized action for the lower
frequency modes. We start from the highest mode $2N$, and the integration
over this mode can be done in a perturbative manner with Gaussian
integration over the variable $q_{2N}\left( \tau \right) $, where $\tau $ is
the imaginary time in the unit of $1/\omega _{z}$. We define a small
parameter $\epsilon =\delta z/z_{0}$, where the length scale $\delta
z=(\hbar /m\omega _{z})^{1/2}$ characterizes the ion oscillation amplitude
for a single ion in a trap with frequency $\omega _{z}$. We consider
renormalization correction to the effective parameters up to the order of $%
\epsilon ^{2}$ (which is the order of $C_{ijkl}$\ term in the action).
Following the standard procedure to calculate the path integral, we find
that after integration of the mode $2N$, the action for the modes $1$ to $%
2N-1$ still takes the form of Eq. (4) up to the order $\epsilon ^{2}$, with
the effective parameters renormalized to
\begin{eqnarray}
\omega _{ij}^{\prime } &=&\omega _{ij}+\epsilon ^{2}\left[ f_{1}\frac{%
C_{i,j,2N,2N}}{2\omega _{2N}}-f_{2}\frac{B_{i,2N,2N}B_{j,2N,2N}}{8\omega
_{2N}^{3}}\right]   \nonumber  \label{5} \\
C_{ijkl}^{\prime } &=&C_{ijkl}+\frac{B_{i,j,2N}B_{k,l,2N}}{4\omega _{2N}^{2}} + O(\epsilon^2)
\nonumber \\
B_{ijk}^{\prime } &=&B_{ijk}+O(\epsilon^2)\\
f_{1} &=&\coth (\frac{\hbar \omega _{z}\omega _{2N}}{2k_{B}T})  \nonumber \\
f_{2} &=&\coth (\frac{\hbar \omega _{z}\omega _{2N}}{2k_{B}T})+\frac{\hbar
\omega _{z}\omega _{2N}}{2k_{B}T}\left[ \sinh (\frac{\hbar \omega _{z}\omega
_{2N}}{2k_{B}T})\right] ^{-2} \nonumber
\end{eqnarray}%
where $\omega _{ij}$ and $\omega _{ij}^{\prime }$ denote the coefficients
before the quadratic term $q_{i}q_{j}$ in the action ($\omega _{ij}=\omega
_{i}^2\delta _{ij}$ in Eq. (\ref{4})), and for the coefficients written as $%
C_{i,j,2N,2N}$ or $B_{i,2N,2N}$, summation over all possible permutations of
the indices are implicitly assumed. After the renormalization, we
re-diagonalize the quadratic term from $\sum_{ij}\omega _{ij}^{\prime
}q_{i}q_{j}$ to $\sum_{i}\omega _{i}^{\prime 2}q_{i}^{\prime 2}$ and make
the corresponding changes to $B_{ijk}^{\prime }$ and $C_{ijkl}^{\prime }$
through change of coordinates from $q_{i}$ to $q_{i}^{\prime }$. With this
step, the action then takes the same form as in Eq. (4), with the mode index
summarizing from $1$ to $2N-1$ and the coefficients renormalized to $\omega
_{i}^{\prime }$, $B_{ijk}^{\prime }$, and $C_{ijkl}^{\prime }$. Then we can
continue with integration of the next highest mode until we finally
integrate out all the modes except for the soft mode 1. The transformation $%
\left( \omega _{i},B_{ijk},C_{ijkl}\right) \rightarrow \left( \omega
_{i}^{\prime },B_{ijk}^{\prime },C_{ijkl}^{\prime }\right) $ defines the RG
flow equations, and after integration of all the modes from mode $N$ to mode
$2$, the last $\omega _{1}^{\prime }$ gives the effective frequency $\omega
_{1eff}$. By numerically solving the RG\ flow equations, the structural
phase transition point can be determined by the criterion $\omega _{1eff}=0$%
. Since the RG flow equations (see Eq. (\ref{5})) depend on the system
temperature $T$, and so does $\omega _{1eff}$, structural phase transition
can be possibly driven solely by temperature under a fixed aspect ratio of
the trap.

The temperature related functions $f_{1}$ and $f_{2}$ can be well
approximated at temperature $T\gg \hbar \omega _{z}\omega _{2N}/k_{B}$
(the latter corresponds to a pretty low temperature compared to Doppler cooling limit) by:
\begin{equation}
f_{1}\simeq \frac{2k_{B}}{\hbar \omega _{z}\omega _{2N}}T,\qquad f_{2}\simeq
2f_{1},
\end{equation}%
so the renormalization correction to $\omega _{1eff}$ is linear in $T$ for a
wide range of temperature. As a result, the critical exponent for
temperature induced linear-to-zigzag phase transition should be $1$, as long
as the critical temperature is above $\hbar \omega _{z}\omega _{2N}/k_{B}$.
The magnitude of the correction to $\omega _{ij}$ at each step is of the
order of $k_{B}T/\left( m\omega _{z}z_{0}^{2}\right) $, which is a small
quantity representing the ratio of system temperature to melting
temperature. It is also worth mentioning that even for zero temperature, the
renormalization correction to $\omega _{1eff}$ is nonzero as $f_{1}=f_{2}=1$
when $T=0$, providing correction from quantum fluctuation to this structural
phase transition.

\begin{figure}[tbp]
\includegraphics[width=0.5\textwidth]{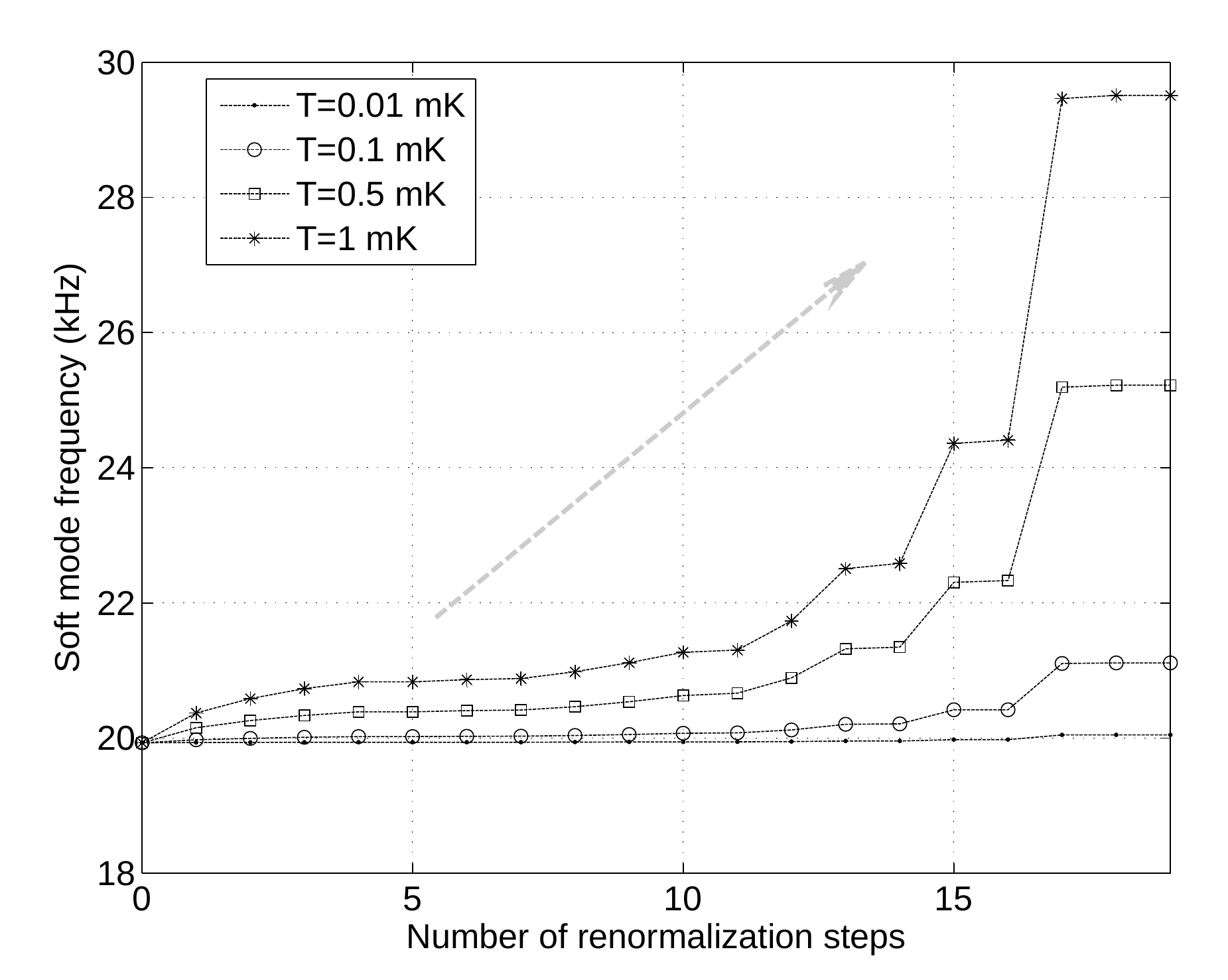}\newline
\caption{(a) Change of the soft mode frequency during the renormalization
process with the aspect ratio $\protect\alpha \equiv \omega_x/\omega_z =4.6$. Different
curves correspond to different temperature, and the number of the renormalization steps represent
the number of high frequency modes that have been integrated out.}
\label{fig1}
\end{figure}

In the following, we carry out some explicit numerical calculation to show
that it is realistic to observe the predicted temperature driven structural
phase transition in the current experimental system. In our calculation, we
take $10$ ions as an example with the mass of ions set as same as $Yb^{+}$
ions. The axial trap frequency is set to $100$ kHz and the aspect ratio $%
\omega _{x}/\omega _{z}$ is chosen around the classical critical value $4.59$
\cite{MF04E}. Temperature is varied on the order from $\mu K$ to $mK$. Fig. %
\ref{fig1} shows the change of soft-mode frequency during the process of
renormalization (the RG\ flow for $\omega _{1}$) at different temperatures.
We find that each renormalization step (integration of one normal mode)
increases slightly the soft mode frequency, and the change after $2N-1$
renormalization steps can be quite significant. The change clearly increases
with the temperature, as the thermal fluctuation of the ion positions
deviate the system from the classical limit where each ion is assumed fixed
at its equilibrium position.

\begin{figure}[tbp]
\includegraphics[width=0.5\textwidth]{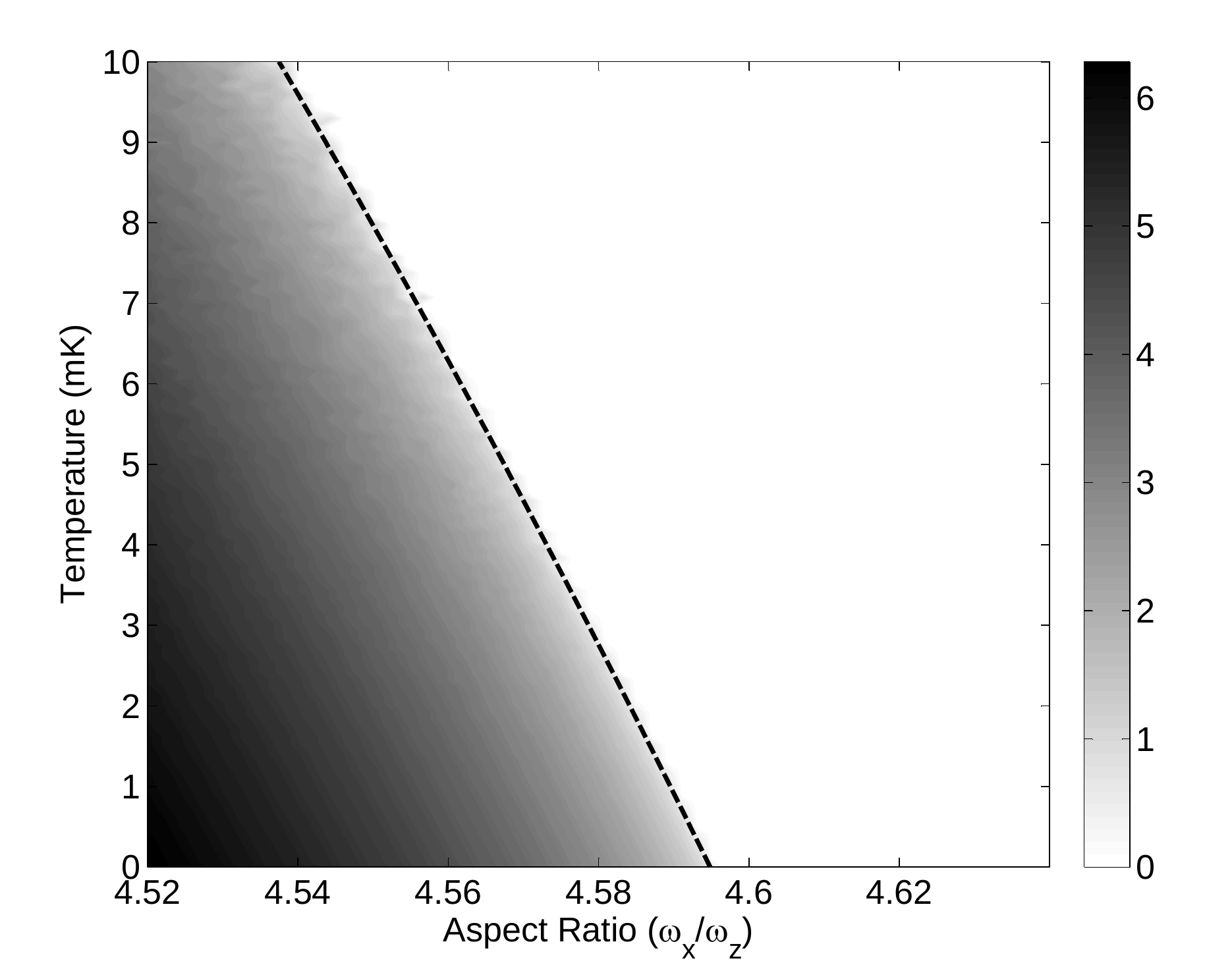}\newline
\caption{The map of the order parameter (with value in unit of $\mu m$) as a function of
temperature and aspect ratio in the linear-to-zigzag phase transition
for $N=10$ ions. The dashed line marks the phase boundary where the order parameter crosses zero.
}
\label{fig2}
\end{figure}

To characterize the phase transition, we calculate the order parameter,
which is taken as the transverse displacement of the zigzag mode (the mean
value of $q_{1}$) for the linear-to-zigzag transition. Fig. \ref{fig2} shows
the value of the order parameter and the corresponding phase diagram as a
function of both temperature and aspect ratio. The phase boundary has a
slope there, which shows that a structure phase transition can be driven
vertically at a fixed aspect ratio solely by change of the system
temperature. From the figure, we also see that the order parameter is more
sensitive to the aspect ratio than to the temperature. Tuning the aspect
ratio by about $1\%$ ($4.59$ to $4.54$ for example) at a fixed temperature
(around $1$ mK) will result in a change of the order parameter by about $5$ $%
\mu m$, while the same change with a fixed aspect ratio around $4.54$
requires one to cool the temperature from $10$ mK to $1$ mK.
\begin{figure}[tbp]
\includegraphics[width=0.4\textwidth]{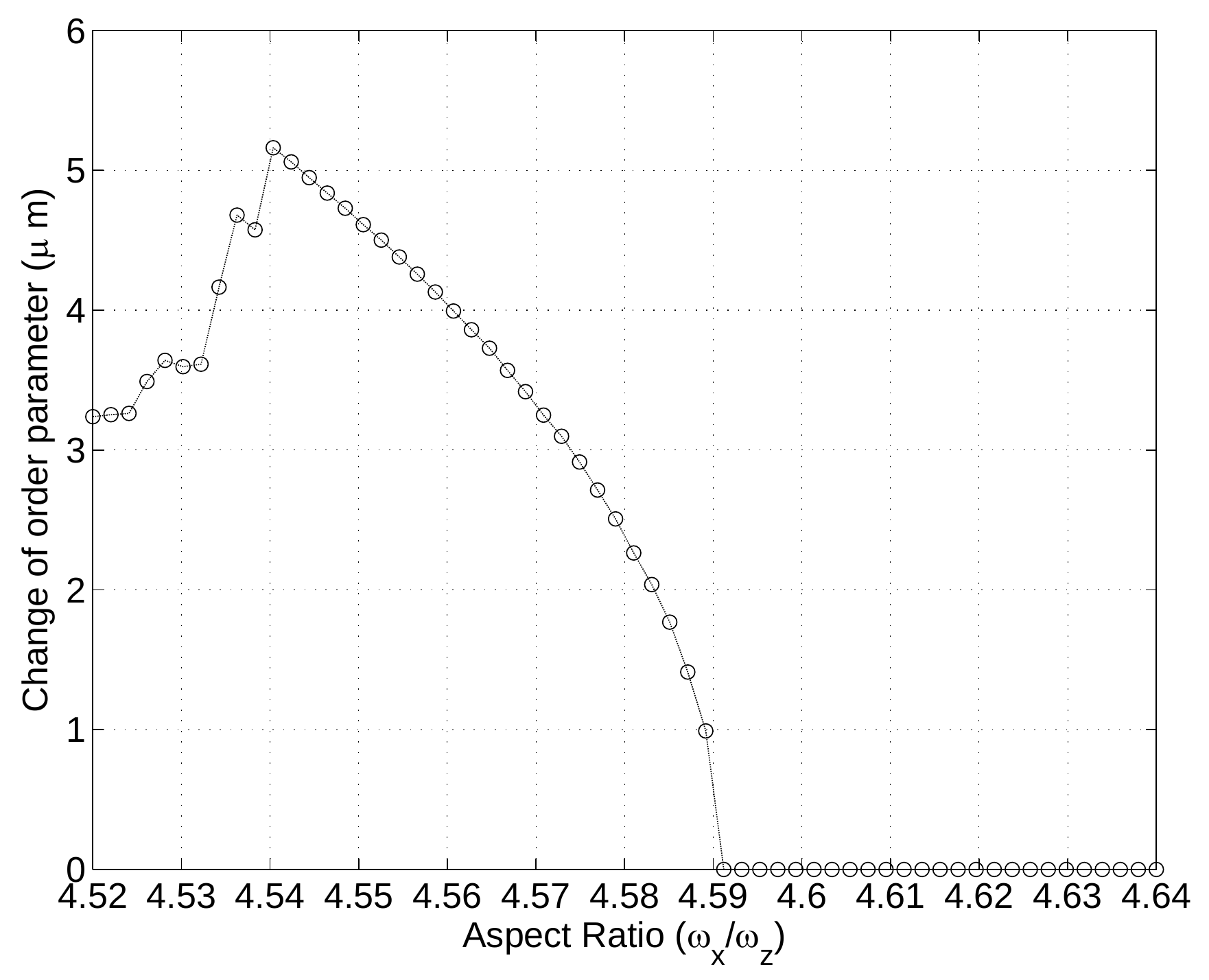}\newline
\caption{The change of value of the order parameter as a function of the aspect ratio when temperature is
cooled from 10mK to 1mK. By tuning the aspect ratio of the confining trap to an optimum value, cooling the
ions can give rise to a change of the order parameter as large as $5\protect\mu m
$, resulting in a fairly noticeable transition from linear to zigzag
pattern. }
\label{fig3}
\end{figure}

Experiment done in Ref. \cite{ESG00} has successfully observed the classcial
linear-to-zigzag phase transition in a trapped ion crystal by changing the
radial trap frequency with an accuracy of $2$ kHz ($0.5\%$ for aspect ratio). With such an accuracy (and probably better
nowadays), one can pick an optimum value for aspect ratio to maximize the
change of order parameter based on the numerical calculation shown in Fig. %
\ref{fig3}. The CCD camera used in Ref. \cite{ESG00} has a resolution of $%
0.3-1$ $\mu m$, which is enough to tell the transition point as the change
of order parameter is apparently larger than $1\mu m$ for a relatively wide range of
aspect ratios (see Fig. 3).

\begin{figure}[h]
\includegraphics[width=0.5\textwidth]{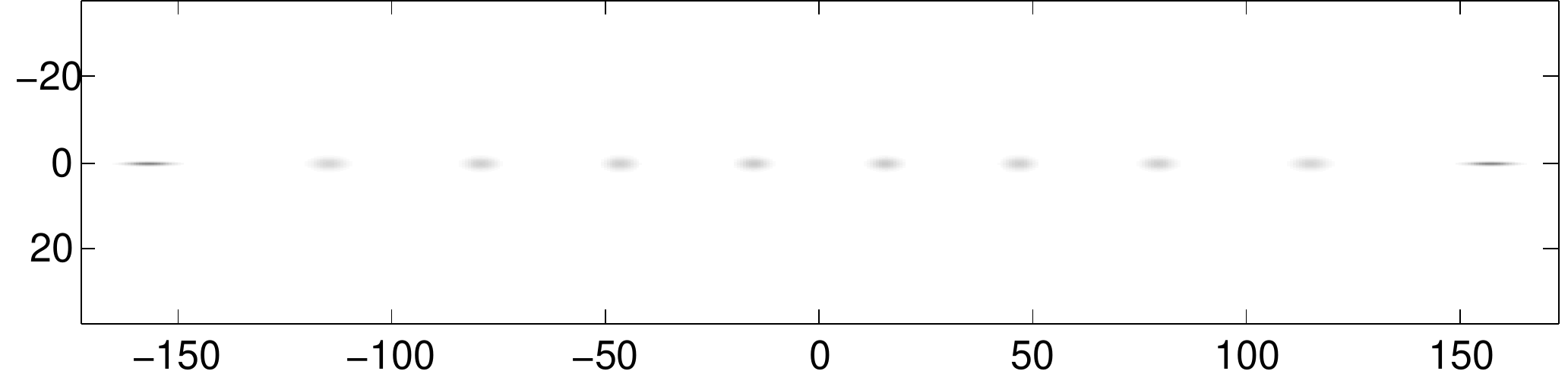}\newline
\includegraphics[width=0.5\textwidth]{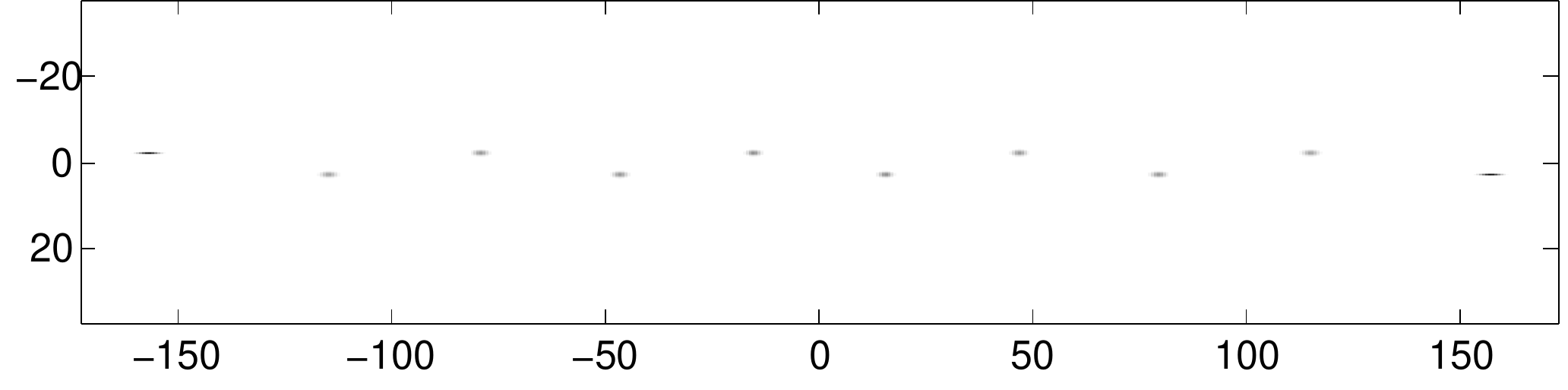}\newline
\caption{Plot of ions' probability density in $x-z$ plane (in unit of $\mu m$ for both
axes) due to thermal fluctuation. The upper figure shows the position and the probability density of $10$ ions
at high temperature (5mK), which characterizes the linear phase.
The lower figure is simulating the ion's position after cooling the
temperature to 1mK, and the zigzag pattern clearly emerges. The aspect ratio
is tuned at about 4.57.}
\label{fig4}
\end{figure}

In real experiments carried out at finite temperature, the thermal
fluctuation of the ions' positions will blur the image of ions. In this
case, we need to calculate whether the image of ions are still sharp enough
to show the temperature driven structural phase transition for the ion
chain. We calculate the thermal fluctuation of ions' axial and transverse
positions, and plot the probability density of the ions' wave-packets above
and below the critical temperature (See Fig. \ref{fig4}), with the aspect
ratio tuned near classical critical value. Here we only demonstrate the case with a few ions (N=10) where the transverse displacement
of all ions can be roughly treated as the same as the order parameter calculated above, but our calculation method works for larger number of ions as well. Our simulation shows that one can clearly observe
the structural phase transition from linear to zigzag pattern, as the
thermal fluctuation of ions' transverse position in the considered temperature range is much smaller than
the change of order parameter across the transition point.

In summary, we have developed a method to characterize the temperature
driven structural phase transition in a trapped ion crystal, taking into
account contributions from both quantum and thermal flucatuation. We use
renormalization group method to calculate the effective soft mode frequency
under finite temperature for a given number of ions and show that the system
have an interesting phase diagram with respect to the system temperature and
aspect ratio. Our predictions can be verified under current experimental
conditions, as shown by our explicit calculations taking account of the
experiemntal imperfections.

This work is supported by the IARPA MUSIQC program, the DARPA OLE Program under ARO Award W911NF0710576,
the ARO MURI program, and the AFOSR MURI program.

\end{document}